\input lanlmac
%
\rightline{EFI-97-28}
\Title{
\rightline{hep-th/9706194}
}
{\vbox{\centerline{Matrix theory and N=(2,1) Strings}}}
\vskip .2in

\centerline{\it Emil Martinec\footnote{*}{Supported 
in part by Dept. of Energy grant DE-FG02-90ER-40560.  } 
}
\bigskip
\centerline{Laboratoire de Physique Th\'eorique et Hautes Energies}
\centerline{Universit\'e Pierre et Marie Curie, Paris VI}
\centerline{4 Place Jussieu, 7252 Paris cedex 05, France}
\smallskip
\centerline{\it and}
\smallskip
\centerline{\footnote{$\dagger$}{Permanent address.}
	Enrico Fermi Inst. and Dept. of Physics}
\centerline{University of Chicago}
\centerline{5640 S. Ellis Ave., Chicago, IL 60637, USA}

\vglue .3cm
\bigskip
 
\noindent
We reinterpret N=(2,1) strings as describing the continuum limit
of matrix theory with {\it all} spatial dimensions compactified.
Thus they may characterize the full set of degrees of freedom needed
to formulate the theory.

\Date{12/96}

\def\journal#1&#2(#3){\unskip, \sl #1\ \bf #2 \rm(19#3) }
\def\andjournal#1&#2(#3){\sl #1~\bf #2 \rm (19#3) }
\def\nextline{\hfil\break}

\def\frac#1#2{{#1\over#2}}

\def\half{\frac12}

\def\inbar{\,\vrule height1.5ex width.4pt depth0pt}
\def\IC{\relax\hbox{$\inbar\kern-.3em{\rm C}$}}
\def\IR{\relax{\rm I\kern-.18em R}}
\def\IP{\relax{\rm I\kern-.18em P}}
\def\Z{{\bf Z}}

%
%
\def\np#1#2#3{Nucl. Phys. {\bf B#1} (#2) #3}

\def\plb#1#2#3{Phys. Lett. {\bf #1B} (#2) #3}

\def\prd#1#2#3{Phys. Rev. {\bf D#1} (#2) #3}

\def\cqg#1#2#3{Class. Quant. Grav. {\bf #1} (#2) #3}
\def\mpl#1#2#3{Mod. Phys. Lett. {\bf #1} (#2) #3}
\def\jmp#1#2#3{J. Math. Phys. {\bf #1} (#2) #3}

\def\nextline{\hfil\break}
\catcode`\@=11
\def\slash#1{\mathord{\mathpalette\c@ncel{#1}}}
\overfullrule=0pt

\def\underrel#1\over#2{\mathrel{\mathop{\kern\z@#1}\limits_{#2}}}

\catcode`\@=12


%

\def\vev#1{\left\langle #1 \right\rangle}


\def\ttilde{{\tilde T}}
\def\stilde{{\tilde S}}


Matrix theory \ref\bfss{T. Banks, W. Fischler, S.  Shenker and L.
Susskind, hep-th/9610043; \prd{55}{1997}{5112}.} 
has been remarkably successful in
capturing the essential ingredients of M-theory --
U-duality groups for compactification on $T^d$ for $d\le4$
(for a summary, see
\ref\uduality{W. Fischler, E. Halyo, A. Rajaraman, and L. Susskind,
hep-th/9703102.}),
the asymptotic Coulomb potential of BPS-saturated sources,
implementation of the `holographic principle'
\nref\holo{L. Susskind, hep-th/9410074; \jmp{36}{1995}{6377}.}%
\refs{\holo,\bfss},
regularization of short-distance supergravity physics
\ref\dkps{M. Douglas, D. Kabat, P. Pouliot, and S. Shenker,
hep-th/9608024; \np{485}{1997}{85}.},
and recovery of string theory when the compactification torus
degenerates appropriately
\nref\motl{L. Motl, hep-th/9701025.}%
\nref\bankseib{T. Banks and N. Seiberg, hep-th/9702187.}%
\nref\dvv{R. Dijkgraaf, E. Verlinde, and H. Verlinde, hep-th/9703030.}%
\refs{\bfss,\motl,\bankseib,\dvv}.

Nevertheless, the formulation is incomplete.
At each stage of compactification, new ingredients are required.
In particular, new degrees of freedom arising from additional
wrapping modes of fluxes make their appearance.
Due to the explosive growth of such modes,
one reaches an impasse in defining matrix theory on $T^6$,
and perhaps even on $T^5$ 
\nref\seiberg{N. Seiberg, hep-th/9705221.}%
\refs{\uduality,\seiberg}.

A seemingly different route to M-theory was initiated by 
D. Kutasov and the author
\nref\kmone{D. Kutasov and E. Martinec, hep-th/9602049;
\np{477}{1996}{652}.}%
\nref\kmo{D. Kutasov, E. Martinec, and M. O'Loughlin, hep-th/9603116;
\np{477}{1996}{675}.}%
\nref\kmtwo{D. Kutasov and E. Martinec, hep-th/9612102;
to appear in {\it Class. Quant. Grav.}.}%
\refs{\kmone-\kmtwo},
who showed that N=(2,1) heterotic strings reproduce the
basic classes of M-theory branes in their target space dynamics.
So far, this approach has received scant attention, perhaps
due to the daunting prospect of using string `field theory'
to describe string worldsheets (although to be fair,
the matrix theory formulation of string worldsheets involves
an equivalent level of complexity -- an infinite tower of
massive modes also decouples from matrix dynamics in the
limit that describes string theory).
In addition, it has been difficult to see how the target string's
interactions might be included in the (2,1) string approach.

Here, it is suggested that matrix theory and N=(2,1) strings
cure one another's ills -- matrix theory provides a natural
framework for interpreting the results of \refs{\kmone-\kmtwo}
as well as for including interactions, while N=(2,1) strings
provide the degrees of freedom needed to define
compactification of all spatial dimensions in matrix theory.
After briefly reviewing the salient facts about N=(2,1) strings,
I describe the proposed correspondence for toroidal
type II vacua, followed by a parallel treatment of
heterotic/type I vacua.  Then the role of U-duality is 
explored, in connection with a generalized Kac-Moody
algebra (GKM) related to $E_{10}$.

\bigskip
\noindent
{\sl N=(2,1) strings in brief}

N=(2,1) heterotic strings couple a modified NSR superstring
in the left-moving chiral sector with the self-dual, integrable
structure of N=2 strings in the right-moving chiral sector.
In a free field representation with
\eqn\fields{\eqalign{
  X^\mu_r\quad,\quad \psi^\mu_r\hskip 1cm&\mu=0,1,2,3\cr
  X^M_\ell \quad,\quad \psi^M_\ell \hskip 1cm& M=0,1,\ldots,11\ ,\cr
}}
the gauge algebra is N=2 local supersymmetry on the right,
with gauge currents $T_r$, $G^\pm_r$, $J_r$; 
and a reducible gauge algebra on the left, generated by N=1
local supersymmetry currents $T_\ell $, $G_\ell $ together with
the null supercurrent $J_\ell =v_M \partial X^M$, $\Psi_\ell =v_M\psi^M$.
The right-moving coordinates have signature $(--++)$,
the left-movers $(--++\ldots +)$.
The current $J_r$ specifies a selfdual two-form $I$
via $J_r=\psi_r^\mu I_{\mu\nu}\psi_r^\nu$.  
The right-moving gauge constraints
are sufficient to eliminate all nonzero modes of that
chirality, leaving only the center-of-mass momentum $p_r$.
We take the time directions noncompact, so that
$p^0_\ell =p^0_r$, $p^1_\ell =p^1_r$; this ensures that the null
constraints imposed by $J_\ell $ bring us to conventional
Lorentzian signature dynamics.
The spatial coordinates will be taken to be compactified
on a torus, so that the spatial momenta lie on a Narain lattice
$\Gamma^{10,2}$.  It will be sufficient for our purpose here
to specialize to 
$\Gamma^{10,2}=({0~1\atop1~0})\oplus({0~1\atop1~0})\oplus
\Gamma_{8}$.

The massless level vertex operators
\eqn\vertexops{\eqalign{
  V_{NS}=&~(e^{-\phi_\ell -\phi^+_r-\phi^-_r})\;\xi_M(p)\psi^M_\ell \;
	e^{ip_\ell \cdot X_\ell +ip_r\cdot X_r}\cr
  V_{R}=&~(e^{-\half\phi_\ell -\phi^+_r-\phi^-_r})\; u_A(p)S_\ell^A
	e^{ip_\ell \cdot X_\ell +ip_r\cdot X_r}\cr
}}
satisfy the constraints 
$v\cdot\xi=p_\ell\cdot\xi=0$, $\xi\sim\xi+\alpha p_\ell+\beta v$;
$\slash v u={\slash p}_\ell u=0$; and
$p_\ell ^2=p_r^2=0$.
The only nontrivial S-matrix elements are the three-point functions
\eqn\threepoint{\eqalign{
  \vev{V_{NS}(1)V_{NS}(2)V_{NS}(3)}=&~
	(\xi_1\cdot\xi_2~ p_2\cdot\xi_3+{\rm cyclic})_\ell \times
	(p_1\cdot I\cdot p_3)_r\cr
  \vev{V_{R}(1)V_{NS}(2)V_{R}(3)}=&~
	({\bar u}_1 {\slash\xi}_2 u_3)_\ell \times
	(p_1\cdot I\cdot p_3)_r\ .\cr
}}
The left-moving structure is that of the 9+1d vector supermultiplet.
If the left- and right-moving momenta were completely independent,
we would interpret the factor $(p_1\cdot I\cdot p_3)_r$
as the structure constants of the gauge group.
In fact the group would be that of symplectic diffeomorphisms
of self-dual null planes in the 2+2 dimensions spanned by $X_r^\mu$,
which is well-known to be the symmetry group of self-dual gravity
\nref\park{Q.-H. Park, \plb{238}{1990}{287}.}%
\nref\ward{R.S. Ward, \cqg{7}{1990}{L217}.}%
\nref\husain{V. Husain, gr-qc/9310003; \cqg{11}{1994}{927}.}%
\refs{\park-\husain}.  Formally, this ${\it SDiff}_2$ group can be thought 
of as SU($\infty$).  Unfortunately, the left- and right-movers
are coupled through the Virasoro constraints $p_\ell^2=p_r^2=0$,
due to the common zero-modes of the time components;
this obstructs the naive interpretation of the massless
sector of the (2,1) string as SU($\infty$) super Yang-Mills (SYM).
Nevertheless, we will keep it in mind as a useful heuristic.
It is likely that this ${\it SDiff}_2$ symmetry is present off-shell
\nref\ov{H. Ooguri and C. Vafa, \np{361}{1991}{469}.}%
\nref\giveon{A. Giveon and A. Shapere, hep-th/9203008;
\np{386}{1992}{43}.}%
\refs{\ov,\giveon}.
Finally, note that \vertexops\ are only the {\it massless} states;
the full spectrum of physical states consists of the 
`Dabholkar-Harvey' states 
\ref\dabharv{A. Dabholkar and J. Harvey, \np{63}{1989}{478}; \nextline
A. Dabholkar, G. Gibbons, J. Harvey, and F. Ruiz Ruiz, \np{340}{1990}{33}.} 
-- right-moving ground states
with arbitrary transverse left-moving excitation, subject only
to the mild constraints of level-matching.  The level density
is exponential \refs{\dabharv,\kmtwo}.

\bigskip
\noindent
{\sl The correspondence with matrix theory}

In \kmone, it was shown that there are generically two nontrivial
decompactification limits of the (2,1) string, depending on
the orientation of the null vector $v_M$.  The simplest choice
places $v$ entirely within the 2+2 dimensions common to both
left- and right-movers.  In this case, the low-energy theory in 
the decompactification limit has the spectrum \kmone\
and dynamics \kmtwo\ of the type IIB D-string in static gauge.
Let us compare this with the corresponding limit of matrix theory.
The matrix theoretic description of M-theory on $T^d$
is given by maximally supersymmetric Yang-Mills (SYM)
on the dual torus $\ttilde^d$
\nref\taylor{W. Taylor, hep-th/9611042.}%
\refs{\bfss,\taylor}.
Perturbative type IIA string theory on $T^{d-1}$ arises from M-theory
upon shrinking a circle to zero size.  In the matrix SYM,
the dual circle decompactifies; the low-energy dynamics is
1+1d SYM \refs{\bfss,\motl,\bankseib,\dvv} --
the spectrum and dynamics of the type IIB D-string!

The second nontrivial decompactification limit of the N=(2,1)
string occurs when the null vector has its spatial component along
one of the purely left-moving coordinates $X_4,...,X_{11}$.
Then the low-energy theory in the limit $R_2,R_3\rightarrow\infty$
has the spectrum of the type IIA D2-brane \kmone.
The Lagrangian agrees to cubic order in the interactions \kmo;
to higher order the situation is not clear, although
general arguments based on symmetries
\nref\mcargese{E. Martinec, lectures at the
1997 Carg\`ese summer school, {\it Strings, Branes, and
Dualities}, May 26--June 14, 1997.}%
\nref\hull{C. Hull, hep-th/9702067.}%
\refs{\kmtwo,\mcargese} (see also \hull) suggest that
the target space dynamics must be physically equivalent to the
D2-brane.  In matrix theory, the decompactification of
a $\ttilde^2$ in the SYM on $\ttilde^d$ yields the type IIB string
\nref\sethisuss{S. Sethi and L. Susskind, hep-th/9702101.}%
\refs{\sethisuss,\bankseib,\uduality};
the low-energy dynamics of the SYM is that of the type IIA
D2-brane -- again there is a direct parallel to the (2,1) string.
Note also that the process stops here; the limit of
a large $\ttilde^3$ is related to that of a small $\ttilde^3$
by U-duality (essentially by T-duality for membranes
\ref\sen{A. Sen, hep-th/9512203; \mpl{A11}{1996}{827}.}).
Small $\ttilde^3$ is {\it decompactification} in M-theory;
one finds a nonperturbative theory on a higher dimensional
spacetime (I will comment briefly on such limits below in a
discussion of U-duality).  
Thus the two distinct weak-coupling IR limits of
matrix theory parallel the two decompactifications of
N=(2,1) string theory.

Matrix theory also provides a neat answer to the question 
of how to build interactions into the target dynamics of
N=(2,1) strings.  Matrix theory suggests \dvv\ that they
should be incorporated through operator insertions 
on a fixed background, rather than topology change of the 
background itself.  It would of course be interesting to
construct such operators for the (2,1) string.

\bigskip
\noindent
{\sl Heterotic/type I vacua}

The next basic class of string vacua are the toroidal compactifications
of heterotic and type I strings.  These are realized as M-theory
on $T^d\times S^1/\Z_2$ 
\ref\horwit{P. Ho\v rava and E. Witten, hep-th/9510209;
\np{460}{1996}{506}.}.
There are several ways to take the weak coupling, perturbative
string limit.  First, one may shrink a single circle to
a size much smaller than the eleven-dimensional Planck scale.
This yields either the $E_8\times E_8$ heterotic or type IA string,
depending on whether this `M-theory circle' is the orbifold
circle $S^1/\Z_2$ or one from $T^d$.  Alternatively,
one may shrink two circles to get a theory more closely related
to IIB strings.  Taking the orbifold circle as one of these,
one obtains the type I/heterotic SO(32) dual pair
(one or the other is weakly coupled depending on whether the
radius $R_{S^1/\Z_2}$ is much smaller/larger than $R_{S^1}$).
Taking both circles from $T^d$, one obtains type IIB with seven-branes --
the F-theory description of the heterotic string on tori.

Some of the matrix-theoretic aspects of this family of vacua
have been worked out
\ref\hetrefs{U. Danielsson and G. Ferretti, hep-th/9610082;\nextline
S. Kachru and E. Silverstein, 9612162;\nextline
L. Motl, hep-th/9612198;\nextline
N. Kim and S.-J. Rey, hep-th/9701139;\nextline
T. Banks and L. Motl, hep-th/9703218;\nextline
D. Lowe, hep-th/9704041;\nextline
S.-J. Rey, hep-th/9704158;\nextline
P. Ho\v rava, hep-th/9705055.}.
%
Matrix quantum mechanics compactified
on $S^1/\Z_2$ is described by 1+1 SYM on a dual circle ${\tilde S}^1$;
the orbifold twist projects the SYM gauge group from U(N) to O(N),
with the transverse coordinates $X^i$ forming a matter multiplet
in the symmetric tensor representation of O(N).
The fermion spectrum surviving the projection (matrices $\theta^a$ in
the adjoint and $\theta^{\dot a}$ in the symmetric tensor)
is anomalous; this anomaly is cancelled by adding 32 fermion fields
$\chi^I$ in the fundamental representation.  Further toroidal
compactification turns some of the $X^i$ into covariant derivatives
on the dual space.  The fact that the original $X^i$ are symmetric
matrices means that the spectrum of the covariant derivatives is
symmetric under reflection -- the dual space is $\ttilde^d/\Z_2$.
To summarize, M-theory on $T^d\times S^1/\Z_2$ is a matrix orbifold
SYM on $\ttilde^d/\Z_2\times {\tilde S}^1$.

Shrinking $S^1/\Z_2$ to get the heterotic string, the dual $\stilde^1$
decompactifies.  The IR description is that of a modified
type I D-string \hetrefs.  
The vacua described above, in which a two-torus
contracts to zero size, correspond to IR limits of 2+1 SYM
on a matrix orientifold $\stilde^1/\Z_2\times\stilde$.

Precisely these sorts of SYM theories arise in the
decompactification limits of a $\Z_2$ orbifold of the
N=(2,1) string (see \kmo\ for details).  The orbifold twist reflects 
$(X^1_r,X^3_r)$ and $(X^1_\ell,X^3_\ell,X^4_\ell,...,X^{11}_\ell)$,
as well as the corresponding superpartners $\psi$.\foot{In a 
convention where 
$J_r=\psi_r\cdot I\cdot\psi_r=\psi^0_r\psi^3_r+\psi^1_r\psi^2_r$.}
The twist acts on the gauge algebra as
\eqn\twist{\eqalign{
  J_r\rightarrow -J_r\quad,&\qquad G^+_r\leftrightarrow G^-_r\cr
  J_\ell\rightarrow -J_\ell\quad,&\qquad 
	\Psi_\ell\rightarrow -\Psi_\ell\ .\cr
}}
One might view this twisted right-moving N=2 algebra as a reduction
of the gauge group of symplectic diffeomorphisms from
SU($\infty$) to SO($\infty$) at the fixed points.
At the massless level, the twisted sector contains 32 massless
chiral fermions, which for example generate the internal 
sector of heterotic strings.

Note that the twist acts on the left-moving `spacetime'
part of the string as though the space coordinates
were on $\ttilde^9/\Z_2\times \stilde^1$; one of the $\Z_2$-twisted 
coordinates is removed by the null projection, leaving
$\ttilde^8/\Z_2\times \stilde^1$.  This would be appropriate to matrix theory
with all transverse dimensions compactified.  Decompactification
yields the type I D-string on $\stilde^1$, if the null vector
points in the directions common to $X_\ell$ and $X_r$;
and an orientifold D2-brane if the spatial component of the
null vector points in the left-moving internal $\Gamma_8$.
These are precisely the ingredients needed to describe
perturbative string limits in the matrix formalism.

\bigskip
\noindent
{\sl U-duality}

Having motivated the link between N=(2,1) strings and
matrix theory, it is appropriate to ask: Where is the 
U-duality group?  A major success of matrix theory is that it
accounts for the entire group of duality symmetries in
high dimensions.  Naively, one might begin by looking to 
classify fluxes of the SYM-like target space theory of (2,1) strings;
identifying them with various Kaluza-Klein modes, wrapped branes, etc.;
and resolving them into duality multiplets.  However, this procedure
will not work so simply in M-theory compactified to such a low dimension as
seems apparent here.  Consider the low-energy theory,
namely 11d supergravity on $T^9$ (or similarly $T^{10}$).
This theory has been analyzed in
\nref\julia{B. Julia, {\it Group Disintegrations}, 
in {\sl Superspace and Supergravity}, S.W. Hawking and M. Ro\v cek, eds.,
Cambridge Univ. Press (1981).}%
\nref\hulltown{C. Hull and P. Townsend, hep-th/9410167;
\np{438}{1995}{109}.}%
\nref\sentwod{A. Sen, hep-th/9503057; \np{447}{1995}{62}.}%
\nref\banksuss{T. Banks and L. Susskind, hep-th/9511193;
\prd{55}{1996}{1677}.}%
\nref\jn{B. Julia and H. Nicolai, hep-th/9608082;
\np{482}{1996}{431};\nextline
H. Nicolai, D. Korotkin, and H. Samtleben, hep-th/9612065.}%
\refs{\julia-\jn}; 
its bosonic sector consists of a scalar field manifold which is a
loop group extension of $E_{8(8)}/SO(16)$, coupled to 2d gravity.
Scalar fields are disordered in two dimensions due to strong IR
fluctuations.  There are no order parameters to orient the
vacuum and provide a fixed background with respect to 
which one can consider the energies of various fluxes.
Similar considerations apply to the $T^{10}$ compactification,
suspected to be related to $E_{10}$ \julia.

Instead, the theory consists of states, which are
wavefunctions on the scalar coset manifold.  The U-duality
group enters by restricting the wavefunctions to be modular
covariant functions under its action\foot{Thanks to J. Harvey 
for remarks clarifying this point.}
(the analogue of restriction from functions
on the upper half-plane $SL(2,\IR)/SO(2)$ to modular functions
under $SL(2,\Z)$).  The wavefunctions might be interchanged
by various U-duality transformations, forming a nontrivial
vector bundle over the moduli space.

Now, in discussions with G. Moore (as reported in \kmtwo),
it was realized that the partition function of the 
toroidally compactified N=(2,0)
string (N=2 string on the right, bosonic string on the left)
has as one-loop partition function the denominator formula
of a generalized Kac-Moody algebra (in that case the Fake Monster
Lie algebra, the natural GKM in 26 dimensions).\foot{For a discussion
of GKM algebras the context of string compactifications, see
\ref\harvmoore{J. Harvey and G. Moore, hep-th/9510182;
\np{463}{1996}{315}; hep-th/9609017.}}
A property of this denominator formula is covariance under
$O(26,2;\Z)$, the Narain group of the (2,0) string.  The
Narain moduli define a character on the root lattice of the GKM.
We further conjectured that the N=(2,1) string realizes the
natural generalization of $E_{10}$ relevant to the superstring.
Again one obtains modular properties under $O(10,2;\Z)$
and a similar interpretation of the Narain moduli.
If this turns out to be the case, it would further cement the connection
between maximally compactified M-theory and N=(2,1) strings,
since this is the expected symmetry group.  At the moment,
our understanding of the quantum N=(2,1) string theory is too
primitive to see its properties under the discrete U-duality group.

The replacement of a moduli space of distinct vacua by modular
functions on moduli space has interesting cosmological implications
\nref\moorehorne{J. Horne and G. Moore, hep-th/940305; 
\np{432}{1994}{109}.}%
\refs{\moorehorne,\banksuss}.
These modular functions are essentially `wave functions of the
universe'.  Their amplitudes in the cusps of the
fundamental domain represent the probabilities of decompactification
to various higher dimensions.  
Given the relation between fields and couplings in string theory,
one also finds a natural probability measure on the space of
coupling constants in these higher-dimensional theories.
One would like to see if four dimensions is somehow preferred.\foot{In
this vein, it was noted in \kmtwo\ that the (2,1) string might serve
as a prototype for an initial condition for cosmology.} 
Higher-dimensional matrix theory compactifications would have to
be recovered in these singular components.  It would be very interesting
to understand the implications for matrix theory on $T^d$
for $d\ge3$.  It is by now well-established that a 
field-theoretic formulation of matrix theory must break down
by the time one reaches five compact dimensions
\refs{\uduality,\seiberg}.  Beyond this it is not known how to proceed.
N=(2,1) strings meet the requirements -- a symplectic diffeomorphism
gauge group structure, and SYM dynamics at low momenta;
with a stringy spectrum of states to regulate the dynamics.


\bigskip
\noindent{\bf Acknowledgements:}  My thanks to
T. Banks,
J. Harvey,
D. Kutasov,
and
G. Moore
for discussions.
A preliminary version of these ideas was presented
at the 1997 Carg\`ese summer school, {\it Strings, Branes, and
Dualities}, May 26--June 14, 1997.


\listrefs
\end